\documentclass[twocolumn,showpacs,preprintnumbers,amsmath,amssymb]{revtex4}
\usepackage{graphicx}% Include figure files
\usepackage{dcolumn}% Align table columns on decimal point
\usepackage{bm}% bold math

\begin{document}

%\draft 
\title{A Fluctuation Probe of Disoriented Chiral Condensates}

\medskip
\author{ B.~Mohanty,$^{1}$ D.P.~Mahapatra$^{1}$ and T.K.~Nayak$^{2}$
}

\medskip

\affiliation{$^{1}$Institute of Physics, Bhubaneswar-751005, India}
\affiliation{$^{2}$Variable Energy Cyclotron Centre, Kolkata-700064, India}

\date{\today}

\begin{abstract}
We show that an event-by-event fluctuation of the ratio of neutral
pions or resulting photons to charged pions can be used as an
effective probe for the formation of disoriented chiral condensates. 
The fact that the neutral pion fraction produced in case of
disoriented chiral condensate formation has a  
characteristic extended non gaussian shape, is shown to be the key
factor which forms the basis of the present analysis.
\end{abstract}


\maketitle

Apart from studying the dynamics of nuclear matter at extreme conditions
of temperature and density, the main goals of relativistic heavy-ion 
collision experiments have been to probe for the possible existence of
quark-gluon plasma (QGP) and chiral symmetry restoration. It is also 
believed that the hot and dense nuclear matter formed in the collision,
while going through the expansion and cooling stage, may relax into a
vacuum state oriented quite differently from the normal ground state.
This results in the formation of what are known as Disoriented Chiral
Condensates (DCC) which finally decay producing an imbalance in pion
production. In such a case the probability distribution of the neutral
pion fraction is very different from what one would expect in the absence
of any DCC production~\cite{bj,ba}. 
It has also been suggested that DCC formation
can be taken as a signal of chiral symmetry restoration at high 
temperature and density~\cite{raj,gavin}. 
Theoretical calculations on lifetime of DCC
gives encouraging results on its detection in actual
experiments~\cite{steele}. 
Several experiments, starting from cosmic ray studies~\cite{cos}, 
hadron-hadron collision~\cite{mini}, to the present day heavy-ion
collision experiments~\cite{wa98_dcc,na49pt} in which 
attempts have been made to look for DCC signal, have either produced
null results or have managed to put an upper limit in DCC production.

Recently fluctuations in the ratio of positive to negative charge and 
baryon to anti-baryon numbers have been shown to be effective probes for the 
possible existence of QGP~~\cite{koch,asakawa,pratt}. Fluctuations in
baryon number or charge asymmetry have been shown to be very much
dependent on whether the system 
goes through a QGP phase or a purely hadronic gas phase. In the present 
letter we show that a similar event-by-event fluctuation study of the
neutral to charged pion ratio can be an effective probe for DCC
production. 

The basic difference between the events with DCC and those without any
DCC formation lies in the probability distribution of the neutral pion
fraction. For events with DCC the probability distribution of neutral
pion fraction, $f$, is given by
\begin{equation}
P(f) = 1/2\sqrt{f} ~~~~{\rm where}~~~
 f = N_{\pi^0}/N_{\pi},
\label{prob}
\end{equation}
$N_{\pi^0}$ and $N_{\pi}$ being the total number of neutral pions and 
the total number of pions respectively. The corresponding 
distribution for non-DCC events is a gaussian with $<f> = 1/3$. It can
easily be seen that for events with DCC there is a strong anti-correlation
in the production of neutral to charged pions. This is the key feature
which forms the basis of the present analysis.

To start with, we define the ratio, $R$ as $N_{\pi^0}$/$N_{ch}$, where
$N_{\pi^0}$ is the number of neutral pions and $N_{ch}$ = $N_{\pi^+}$
+ $N_{\pi^-}$, is the multiplicity of charged pions. The
fluctuation in the ratio $R$  is given as

\begin{eqnarray}
D = \frac{<\delta R^2>}{{<{R}>}^2}
= \frac{<\delta N_{\pi^{0}}^{2}>}{{<{N_{\pi^{0}}}>}^2}
+ \frac{<\delta N_{{ch}}^{2}>}{{<{N_{{ch}}}>}^2} \nonumber \\
-2 \frac{<\delta{N_{{ch}}}\delta{N_{\pi^{0}}}>}{{<{N_{{ch}}}>}
{{<{N_{\pi^{0}}}}>}}
\label{fluc_def}
\end{eqnarray}
where the average ($< \cdots >$) is over events with \\
$<\delta R^2> = <R^{2}> - <R>^{2}$.

Since the total probability is always $1$,
knowing $N_{\pi^0}$ = $f~N_{\pi}$, one can write
$N_{ch}$ = $(1-f)~N_{\pi}$. Following simple statistical analysis, we 
can show

\begin{equation}
\frac{\delta{N_{\pi^{0}}}}{{<{N_{\pi^{0}}}>}} = 
\frac{\delta{N_{\pi}}}{{<{N_{\pi}}>}} +
\frac{\delta{f}}{{<{f}>}} 
\label{n_der}
\end{equation}
and
\begin{equation}
\frac{\delta{N_{{ch}}}}{{<{N_{{ch}}}>}} = 
\frac{\delta{N_{\pi}}}{{<{N_{\pi}}>}} -
\frac{\delta{f}}{{(1-<{f}>)}} 
\label{c_der}
\end{equation}  

By squaring Eq.~(\ref{n_der}) and Eq.~(\ref{c_der}) one can get the
contributions to the  
terms  
${<\delta N_{\pi^{0}}^{2}>}/{{<{N_{\pi^{0}}}>}^2}$ and
${<\delta N_{{ch}}^{2}>}/{{<{N_{{ch}}}>}^2}$ in
Eq.~(\ref{fluc_def}). While by multiplying Eq.~(\ref{n_der}) and
Eq.~(\ref{c_der}), the contribution to 
the cross term 
${<\delta{N_{{ch}}}\delta{N_{\pi^{0}}}>}/{{<{N_{{ch}}}>}
{{<{N_{\pi^{0}}}}>}}$ in Eq.(~\ref{fluc_def}) is obtained. 
So the fluctuation in ratio $R = N_{\pi^0}/N_{ch}$, as given by
Eq.~(\ref{fluc_def}) becomes

\begin{equation}
D =
\frac{<\delta{f}^2>}{{{{<{f}}>}^2}{(1-<{f}>)}^2}
\label{fluc_final}
\end{equation} 

This shows that the fluctuation in the ratio, $R$, is
essentially dominated by the fluctuation in the neutral pion 
fraction whose probability distribution for DCC is given in 
Eq.~(\ref{prob}). Now let us see, what is the value of 
$D = {<\delta R^2>}/{{<{R}>}^2}$
for non-DCC and DCC cases respectively. \\

\noindent{ \bf \it Non-DCC case. - } It is known from $pp$ experiments
~\cite{pp_exp} that the produced pions have their charge states 
partitioned binomially with the mean of $f$ at $1/3$. The fluctuation 
in $f$ is inversely proportional to the total number as given by
${<\delta{f}^2>}$  $\sim$ 2/9$N_{\pi}$.
Then one can write
\begin{equation}
D_{non-DCC}= \frac{1}{N_{\pi}{{{<{f}>}^2}{(1-<{f}>)}^2}}
                     \sim 4.5/N_{\pi}
\label{nodcc}
\end{equation}

\noindent{\bf \it  DCC case. - } For the DCC case, the probability
distribution is given by Eq.~(\ref{prob}). 
Using this, one can easily see that $<f> = 1/3$, as in the non-DCC
case, while $<\delta{f}^2>
= 4/45$. Substituting these values in Eq.~(\ref{fluc_final}), we find,
\begin{equation}
D_{DCC} = 1.8
\label{withdcc}
\end{equation}
For a typical case, where the total pion multiplicity, $N_{\pi}$,  
in the experiment is 300, we find for the non-DCC case the above
fluctuation is only about 0.015. This is way below the value 1.8 obtained
for DCC. \\

In a typical reaction DCC may or may not be formed and if formed
there could even be background pions coming from non-DCC sources.
There could also be multiple DCC domains formed. The effects of 
these are discussed below.

\begin{figure}
\begin{center}
\includegraphics[scale=0.4]{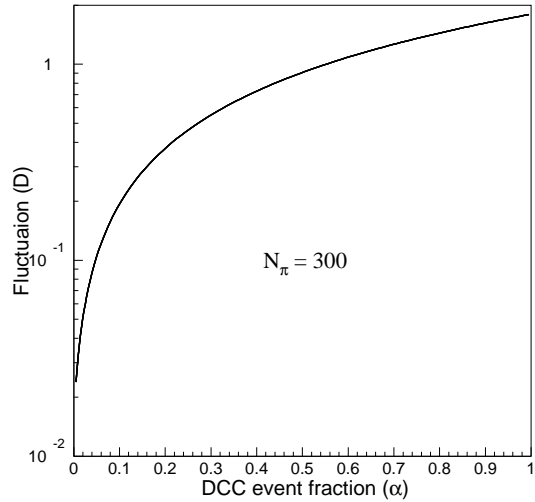}
\caption {\label{dcc_fluc}
Variation of fluctuation (D) with the fraction of DCC events ($\alpha$).
The plot is obtained for a average detected pion multiplicity of 300.
}
\end{center}
\end{figure}

\noindent {\bf \it Ensemble of DCC and non-DCC events. -}
In reality, DCC formation depends on
several factors~\cite{dcc_prob}. That means not all events in a
collision are favourable for DCC production. If the sample of events
is a mixture of DCC and non-DCC events, the signal should 
depend on the respective fractions. To a first approximation, 
the corresponding fluctuation would be a linear combination of
results shown in Eq.~(\ref{nodcc}) and ~(\ref{withdcc}) as given by
\begin{equation}
 D = \alpha D_{DCC} + (1-\alpha) D_{non-DCC}
\label{mix}
\end{equation}
where $\alpha$ is the DCC event fraction. 
Fig.~\ref{dcc_fluc} shows the variation of fluctuation, $D$,
with the fraction of DCC events ($\alpha$), obtained from 
Eq.~(\ref{mix}), for a typical case of average total pion
multiplicity of 300.

\noindent{\bf \it Multiple DCC domains. -} 
The fluctuation of $D_{DCC}$ as given in Eq.~(\ref{mix}) is valid for
pions emitted
from a single region or domain of DCC. It is possible that in an event
multiple domains of DCC are formed. In such a case the 
total probability distribution of the neutral pion fraction is the 
average value of $P(f)$ over all of the domains. This
can be written as
\begin{eqnarray}
      P_{m}(f) = \int df_{1} \cdots df_{m} \delta( f -
      \frac{f_{1}+ \cdots +f_{m}}{m}) \nonumber \\
 P_{1}(f_{1}) \cdots P_{m}(f_{m})
\label{mdom}
\end{eqnarray}
where, $m$ is the number of domains and for the non-DCC case $P(f) \sim
\delta(f-1/3)$. For a case with two domains Eq.~(\ref{mdom}) can be 
evaluated analytically to yield a value 
$\pi/2$  if $f<1/2$ and $\pi/2-2arccos(1/2\sqrt(f))$ for $f>1/2$.
For higher number of domains one has to compute it numerically. It
can be shown that the resultant probability distribution approaches
a gaussian centered at 1/3 with the
standard deviation $\sim$ $1/\sqrt{m}$.
This means ${<\delta{f}^2>}/{{{<{f}}>}^2}$  $\sim 1/ m$, so
that the fluctuation, $D$, reduces as the number of domains 
increases in an event. However, by carrying out this analysis
by dividing the $\eta-\phi$ phase space to appropriate bins
this effect can be reduced.

\noindent{\bf \it Events with pions from DCC as well as non-DCC
  sources. -}
In a given event pions can originate both from a DCC source and
from other non-DCC sources even in a DCC event.
In such a case the probability distribution for the neutral 
pion fraction, $P(f)$, can be evaluated following Eq.~(\ref{mdom}) as

\begin{eqnarray}
P(f) = \int~df_{DCC}~df_{non-DCC}~ \nonumber \\
P(f_{DCC})~P(f_{non-DCC}) \nonumber \\ 
 \delta(f -\beta f_{DCC}-(1-\beta)f_{non-DCC})
\label{new_prob}
\end{eqnarray}

\begin{figure}
\begin{center}
\includegraphics[scale=0.4]{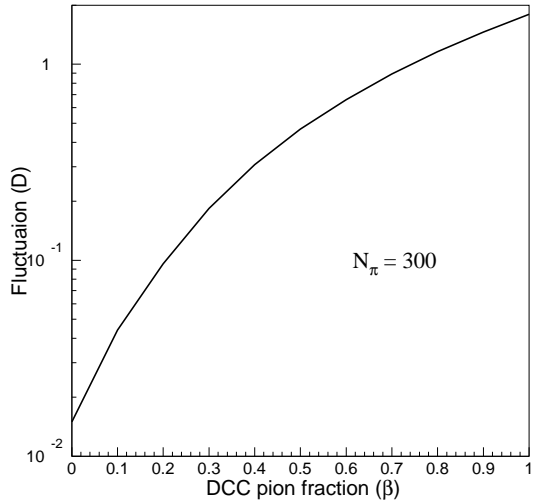}
\caption {\label{dcc_pion}
Variation of fluctuation (D) as a function of the pion fraction 
($\beta$) from a DCC origin.
The plot is obtained for an average pion multiplicity of 300.
}
\end{center}
\end{figure}

\noindent{where $\beta$ corresponds to fraction of DCC pions out of a total 
number of pions $N_{\pi}$, produced in the event. $f_{DCC}$ and
$f_{non-DCC}$ correspond to the neutral pion fraction for
pions from DCC and non-DCC sources in the event, respectively.} 
With the presence of non-DCC
pions, one can see, the neutral pion fraction can no longer start
from zero nor can reach the full value of unity, the range depending
upon the fraction of non-DCC pions present in the sample. 
For various values of $\beta$, and total number of non-DCC pions,
$(1-\beta)N_{\pi}$, the above equation can be evaluated easily
to obtain the final $f$ distribution from which one can get the
resultant fluctuation, $D$, using Eq.~(\ref{fluc_final}). 
The results obtained for an average total number of 300 pions
in an event, are shown in Fig.~\ref{dcc_pion}. One can see that the 
fluctuation $D$, decreases with decrease in
the fraction of DCC pions. 

\noindent {\bf \it Neutral pion decay. -}Neutral pions are not actually 
detected. They are rather 
reconstructed from photons detected using electromagnetic
calorimeters. Basically one detects the decay photons ($\pi^{0}$
$\longrightarrow$ $2\gamma$).
For such a case one can take $N_{\pi^{0}} = 2N_{\gamma}$.
However we find that this substitution leaves the 
expression in Eq.~(\ref{fluc_def}) unchanged. That is
\begin{equation}
D = D_{\gamma}
\end{equation}
where $D_{\gamma}$ corresponds to the fluctuation in the ratio 
$N_{\gamma}/N_{ch}$. That makes this observable actually measurable,
event-by-event, in an experiment and the conclusions derived 
above for the
$N_{\pi^{0}}$/$N_{{ch}}$ 
hold true for $N_{\gamma}$/$N_{{ch}}$. 
In an actual experiment, such as in STAR at RHIC, one can measure 
$N_{\gamma}$ and $N_{{ch}}$
and look for the signal of DCC using this method. 

We know that the measurement of any observable in an 
experiment is affected by the
efficiency of the detector. The effect of this can be seen easily
through the following calculation. Consider the efficiency of
detecting photons is $\epsilon_{1}$ and that for charged pions is
$\epsilon_{2}$. Then we have, $N^{exp}_{\gamma}$ = $\epsilon_{1}
N_{\gamma}$ and $N_{{ch}}^{exp}$ = $\epsilon_{2} N_{{ch}}$.
Following the simple statistical analysis as used for arriving at
Eq.~(\ref{n_der}), we get,

\begin{eqnarray}
D^{exp} 
= D + 
\frac{<\delta \epsilon_{1}^{2}>}{{<{\epsilon_{1}}>}^2} +  
\frac{<\delta \epsilon_{2}^{2}>}{{<{\epsilon_{2}}>}^2}  
\label{exp}
\end{eqnarray}

\noindent{provided we assume that the efficiency of detection 
is independent of
multiplicity. This is not a bad assumption if the analysis is carried
out over a narrow bin in centrality.}

With detectors for measuring photon and charged particle
multiplicity in a common coverage of the $\eta-\phi$ phase space
as in WA98~\cite{wa98_dcc} in CERN SPS, STAR in RHIC and ALICE 
at LHC, the present method of analysis is expected to be very 
useful regarding the search for DCC signal. The observed photon 
multiplicity, $N_{\gamma-{\mathrm like}}$, in such a situation
is usually contaminated with some charged particles that mimic
photons. However, the true photon multiplicity, $N_{\gamma}$ can
be extracted from the measured $N_{\gamma-{\mathrm like}}$ value 
after correcting for the efficiency of photon counting and
the purity of photon sample. This is done using the relation
~\cite{wa98-9}, 
\begin{equation}
      N_{\gamma} =
      \frac{p_\gamma}{\epsilon_{\gamma}}~N_{\gamma-{\mathrm like}}  
\end{equation}
where, $\epsilon_{\gamma}$ is the photon counting efficiency and 
$p_\gamma$ is the purity of the photon sample obtained from detector
simulations. This will modify the expression for 
fluctuation given in Eq.~(\ref{exp}). If we assume that the purity and
efficiency are not dependent on multiplicity and independent of each 
other, we get
\begin{equation}
D^{exp}
=  
D +
\frac{<\delta \epsilon_{\gamma}^{2}>}{{<{\epsilon_{\gamma}}>}^2} +  
\frac{<\delta \epsilon_{2}^{2}>}{{<{\epsilon_{2}}>}^2}  +
\frac{<\delta p_{\gamma}^{2}>}{{<{p_{\gamma}}>}^2} 
\end{equation}

\noindent{\bf \it Sensitivity to DCC signal. -}
Let us consider an experiment where the average total detected 
pion multiplicity is $300$. Taking typical values of relative
fluctuation ($\sigma/mean$) in efficiency and purity, which are
of the same order, to be
$\sim 3\%$~\cite{wa98_fluc}, we have
${<\delta \epsilon^{2}>}/{{<{\epsilon}>}^2}$ and  
${<\delta p_{\gamma}^{2}>}/{{<{p_{\gamma}}>}^2}$ both to 
be about $0.0009$. Together they introduce an error
$0.0027$ in the value of fluctuation, $D_{non-DCC}$, which now
has a value $0.015~\pm 0.0027$. 

From Fig.~\ref{dcc_fluc} we find that corresponding to $D \sim 0.02$
which is the case for pions from non-DCC events for a typical
experiment as described here, the value of $\alpha~=~0.002$. This means
that the analysis is sensitive if the fraction of DCC events in a given
sample of events is above $0.2\%$, provided all the pions in the
sample of DCC events are of DCC origin.
From Fig.~\ref{dcc_pion} we find a value of $\beta=0.025$ 
corresponding to $D \sim 0.02$. This means that in a sample of DCC
events, the method is sensitive if more than $2.5\%$ of the pions are of
DCC origin. 

\begin{figure}
\begin{center}
\includegraphics[scale=0.4]{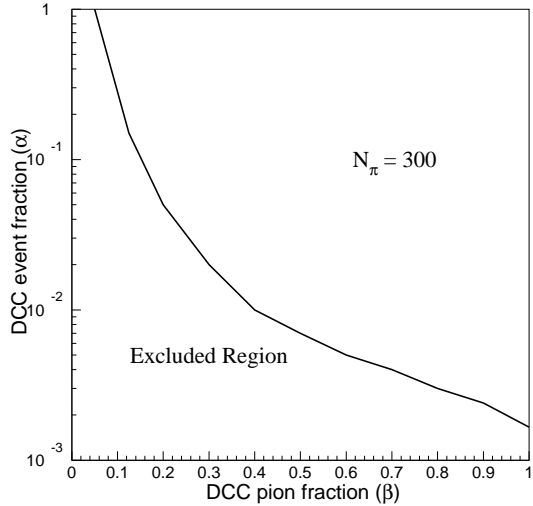}
\caption {\label{dcc_sen}
The sensitivity of a typical experiment in terms of the fraction 
of DCC type events, $\alpha$, for a given fraction of pions coming
from DCC origin, $\beta$, corresponding to an average detected
pion multiplicity of 300.
}
\end{center}
\end{figure}

Combining the results from Fig.~\ref{dcc_fluc} and
Fig.~\ref{dcc_pion} one can obtain a sensitivity curve 
for such a typical experiment.
This is shown in Fig.~\ref{dcc_sen}. It shows, the sensitivity 
of the method in a typical experiment to the fraction of DCC pions,
$\beta $, in an  
event vs. the fraction of DCC events, $\alpha $. 
The contour is drawn corresponding to a value of $D \sim 0.02$.
The excluded region, which is to the left 
of the curve, indicates the region which cannot be probed by such 
an experiment. For example, the experiment is sensitive to 
DCC with $20\%$ of the pions coming from a DCC origin, provided 
there are more than $4$ events in every $100$ events which are 
DCC-type. By improving on the photon detection and reducing the
fluctuation in efficiency and purity of measured photon sample, the
sensitivity of the method can be improved. In experiments at RHIC and
LHC, where the number of emitted pions are significantly large, the
fluctuation in the ratio of photons to charged particles for
non-DCC sources are going to be quite small. Thus the present method will
be quite sensitive for detecting signals of DCC at these energies.

The effect of acceptance on ratio fluctuations such as ours, 
have been discussed in Ref.~\cite{koch}. As has been argued there,
the results are expected to be independent of acceptance. Some
more general discussion on effect of acceptance on multiplicity 
fluctuations can be found in Ref~\cite{accp}.

In summary, we have shown that detection of DCC is possible 
through an event-by-event fluctuation study of the ratio 
of multiplicities of neutral to charged pions. For a case, with
DCC production, the above fluctuation has a value close to $2$
while for a non-DCC case the same turns out to be much smaller,
inversely proportional to the pion multiplicity. However in an 
actual experiment, event-by-event measurement of neutral pion
multiplicity may not be feasible. Instead one detects photons 
coming from its decay. We have shown that the conclusions drawn 
from the fluctuation in $N_{\pi^0}$/$N_{ch}$ 
remain unchanged for the fluctuation in the ratio of 
photon and charged pion multiplicities ($N_{\gamma}$/$N_{ch}$). 
The analysis has been extended to include pions from both DCC
and non-DCC sources in the same event. We have discussed other 
detector related effects on this and demonstrated that the proposed 
fluctuation analysis can easily be applied to experimental data 
regarding DCC search. 

The authors wish to thank S. C. Phatak, Institute of
Physics, and N. Barik, Department of Physics, Utkal University,
Bhubaneswar for some very helpful suggestions and discussions.


\medskip

\begin{thebibliography}{99}

\bibitem{bj} J.D.~Bjorken, Int. J. Mod. Phys. {\bf A7}, 4189
  (1992).

\bibitem{ba}     J. -P. Blaizot and A. Krzywicki,
                    Phys. Rev. {\bf D46}, 246 (1992).


\bibitem{raj} K.~Rajagopal and F.~Wilczek, Nucl. Phys. {\bf B399}, 395 
  (1993); Nucl. Phys. {\bf B404}, 577 (1993).


\bibitem{gavin}    S. Gavin, A. Gocksch and R.D. Pisarski,
                      Phys. Rev. Lett. {\bf 72}, 2143 (1994).


\bibitem{steele}      J.V. Steele and V. Koch,
                      Phys. Rev. Lett. {\bf 81}, 4096 (1998).


\bibitem{cos}   C.R.A. Augusto et al., Phys. Rev. {\bf D59}, 054001 (1999).


\bibitem{mini}   T.C. Brooks et al., (Minimax Collaboration),
                 Phys. Rev. {\bf D61}, 032003 (2000).

\bibitem{wa98_dcc}    M.M.~Aggarwal et al., (WA98 Collaboration), Phys. Lett.
                      {\bf B420}, 169 (1998);
                     Phys. Rev. {\bf C58}, 011901(R) (2001).

\bibitem{na49pt}    H. Appelshauser et al., (NA49 Collaboration),
                      Phys. Lett. {\bf B459}, 679 (1999).


\bibitem{koch}   S.~Jeon, V.~Koch,
                      Phys. Rev. Lett. {\bf 85}, 2076 (2000).



\bibitem{asakawa}   M.~Asakawa, U.~Heinz, B.~Muller,
                      Phys. Rev. Lett. {\bf 85}, 2072 (2000).

\bibitem{pratt}   S.A.~Bass, P.~Danielewicz, S.~Pratt,
                      Phys. Rev. Lett. {\bf 85}, 2689 (2000).


\bibitem{pp_exp}  P.~Grassberger, H.I.~Miettinen, 
                  Nucl. Phys. {\bf B89}, 109 (1975).

\bibitem{dcc_prob}      A. Krzywicki and J. Serreau,
                    Phys. Lett. {\bf B448}, 257 (1999).


\bibitem{wa98-9}    M.M.~Aggarwal et al., (WA98 Collaboration),
                       Phys. Lett. {\bf B458}, 422 (1999).

\bibitem{wa98_fluc}    M.M.~Aggarwal et al., (WA98 Collaboration), 
                       Submitted to Physical Review C,
                       eprint: nucl/ex-0108029.



\bibitem{accp}  D.P.~Mahapatra, B.~Mohanty, S.C.~Phatak,
                    e-print:nucl-ex/0108011;
                    To be published in Int. J. Mod. Phys. {\bf A}.


\end{thebibliography}
\end{document}